\documentclass[12pt,superscriptaddress]{revtex4}
\usepackage{graphicx}

\newcommand{\be}{\begin{equation}}
\newcommand{\ee}{\end{equation}}
\newcommand{\bea}{\begin{eqnarray}}
\newcommand{\eea}{\end{eqnarray}}

\begin{document}
\title{Mode-coupling and the pygmy dipole resonance in a relativistic two-phonon model}
\author{Elena Litvinova}
\affiliation{GSI Helmholtzzentrum f\"{u}r Schwerionenforschung,
64291 Darmstadt, Germany}
\affiliation{Institut f\"{u}r Theoretische Physik,
Goethe-Universit\"{a}t, 60438 Frankfurt am Main, Germany}
\affiliation{Institute of Physics and Power Engineering, 249033
Obninsk, Russia}
\author{Peter Ring}
\affiliation{Physik-Department der Technischen Universit\"at
M\"unchen, D-85748 Garching, Germany}
\author{Victor Tselyaev}
\affiliation{Nuclear Physics Department, St. Petersburg State
University, 198504 St. Petersburg, Russia}

\date{\today}

\begin{abstract}
A two-phonon version of the relativistic quasiparticle time blocking
approximation (RQTBA-2) represents a new class of many-body models
for nuclear structure calculations based on the covariant energy
density functional. As a fully consistent extension of the
relativistic quasiparticle random phase approximation (RQRPA), the
two-phonon RQTBA implies a fragmentation of nuclear states over
two-quasiparticle and two-phonon configurations. This leads, in
particular, to  a splitting-out of the lowest 1$^-$ state as a
member of the two-phonon $[2^+\otimes3^-]$ quintuplet from the RQRPA
pygmy dipole mode, thus establishing a physical mixing between these
three modes. The inclusion of the two-phonon configurations in the
model space allows to describe the positions and the reduced
transition probabilities of the lowest 1$^-$ states in isotopes
$^{116,120}$Sn as well as the low-energy fraction of the dipole
strength without any adjustment procedures. The model is also
applied to the low-lying dipole strength in neutron-rich
$^{68,70,72}$Ni isotopes. Recent experimental data for $^{68}$Ni are
reproduced fairly well.
\end{abstract}

\maketitle


The theoretical description of nuclear low-lying dipole strength
remains among the most important problems in nuclear structure and
nuclear astrophysics~\cite{PVKC.07}. Measurements of the dipole
strength by means of high resolution nuclear resonance
fluorescence~\cite{Rye.02,ZVB.02,SBB.06,SRB.07,SRT.08,SFH.08}
resolve the fine structure of the spectra below the neutron
threshold. Unique spectroscopic information about neutron-rich
medium-mass and heavy nuclei have been obtained in recent
experiments with Coulomb dissociation~\cite{Adr.05,KAB.07} and
virtual photon scattering~\cite{WBC.09}. This offers exciting
opportunities for microscopic nuclear structure models to describe
the fine structure of the dipole spectra below the neutron threshold
and to help in analyzing the experimental data.

Measurements of the low-lying dipole strength above and below the
neutron threshold are performed with different nuclear reactions,
which have reduced sensitivity in the area around the threshold.
Therefore, a correct comparison of the calculated pygmy strength
with the data is still problematic. The collectivity of the pygmy
mode is another subject of discussions. Self-consistent relativistic
QRPA calculations~\cite{PRN.03} produce a highly collective pygmy
mode, in contrast to the results of the non-relativistic
approaches~\cite{TSG.07,TL.08}. However, in Ref.~\cite{TE.06} it is
pointed out that the collectivity of the pygmy mode is restored
within non-relativistic QRPA calculations with Skyrme forces when a
fully self-consistent scheme is employed.

Since the pygmy dipole mode has essentially surface nature, it mixes
with other surface modes, especially with low-lying ones. This
supposition has been confirmed by explicit RQTBA
calculations~\cite{LRT.08,LRTL.09} for tin and nickel isotopes and
N=50 isotones. It has been found that the pygmy mode, arising in
RQRPA as a single state or as very few low lying dipole states with
isoscalar character, is strongly fragmented over many states in a
broad energy region due to the coupling to phonons. As a result,
some fraction of the strength is located well below the original
position of the RQRPA pygmy mode. Compared to existing data, the
RQTBA describes the total fraction of the dipole strength below the
neutron threshold very reasonably~\cite{LRT.08,LRTL.09}. However, in
order to account for the fine structure of the spectrum, more
correlations should be included in the microscopic model.
%
%
In particular, it has to reproduce the lowest dipole state in
vibrational nuclei, which is identified as a member of the
quintuplet $2^+_1\otimes3^-_1$, as it was predicted in
Refs.~\cite{Lip.71,VK.71} and observed in spectra of spherical
nuclei~\cite{WRZ.96,Bry.99,Pys.06,OEN.07}.

In the present work, the two-phonon version of the quasiparticle
time blocking approximation (QTBA) proposed in Ref.~\cite{Tse.07} is
employed to introduce correlations between the two quasiparticles
within the 2q$\otimes$phonon configurations of the RQTBA. Therefore,
a fragmentation of the nuclear states over the two-phonon
configurations appears in the excitation spectra in addition to the
spreading over the two-quasiparticle states.
%
%
\begin{figure*}[ptb]
\begin{center}
\includegraphics*[scale=0.8]{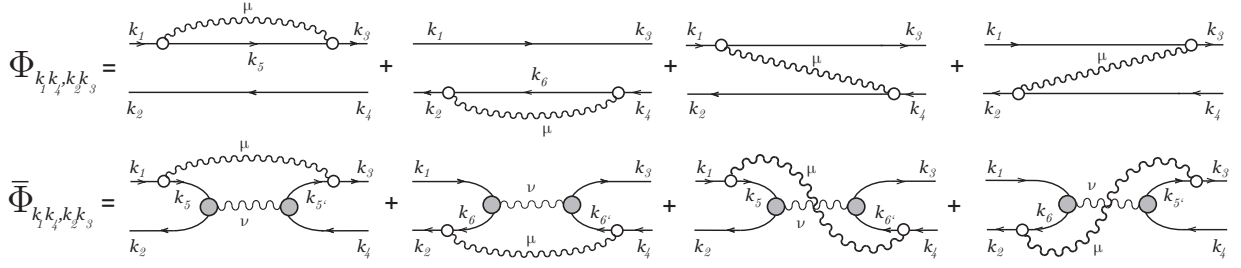}
\end{center}
\caption{The correspondence between the 2q$\otimes$phonon amplitude
$\Phi$ of the conventional phonon coupling model and the two-phonon
amplitude $\bar{\Phi}$ of the two-phonon model in a diagrammatic
representation. Solid lines with arrows and latin indices denote the
single-quasiparticle nucleonic propagators, wavy curves with greek
indices the phonon propagators, empty circles represent phonon
vertices, and grey circles together with the two nucleonic lines
denote the RQRPA transition densities (see text).}
\label{f1}%
\end{figure*}
As well as in the conventional RQTBA, excitations in Fermi-systems
with even particle number are described by the Bethe-Salpeter
equation (BSE) containing both static and energy-dependent residual
interactions. This equation for the nuclear response function
$R(\omega)$ in the doubled quasiparticle space reads:
\be
R_{k_{1}k_{4},k_{2}k_{3}}^{\eta\eta^{\prime}}(\omega)=\tilde{R}_{k_{1}k_{2}%
}^{(0)\eta}(\omega)\delta_{k_{1}k_{3}}\delta_{k_{2}k_{4}}\delta^{\eta
\eta^{\prime}}+\tilde{R}_{k_{1}k_{2}}^{(0)\eta}(\omega)
\sum\limits_{k_{5}k_{6}\eta^{\prime\prime}}{\bar{W}}_{k_{1}k_{6}%
,k_{2}k_{5}}^{\eta\eta^{\prime\prime}}(\omega)R_{k_{5}k_{4},k_{6}k_{3}}%
^{\eta^{\prime\prime}\eta^{\prime}}(\omega),
\label{respdir}%
\ee
where the indices $k_i$ run over single-particle quantum numbers
including states in the Dirac sea and the indices $\eta,
\eta^{\prime}, \eta^{\prime\prime}$ numerate forward $(+)$ and
backward $(-)$ components in the doubled quasiparticle space. The
quantity $\tilde{R}^{(0)\eta}_{k_1k_2}(\omega) = 1/(\eta\omega -
E_{k_1} - E_{k_2})$ describes the free propagation of two
quasiparticles with their Bogoliubov's energies $E_{k_1}$ and
$E_{k_2}$ in the mean field between subsequent interactions with the
amplitude
\be
{\bar{W}}_{k_{1}k_{4},k_{2}k_{3}}^{\eta\eta^{\prime}}(\omega)=\tilde{V}%
_{k_{1}k_{4},k_{2}k_{3}}^{\eta\eta^{\prime}}
+ \Bigl[\Phi_{k_{1}k_{4},k_{2}%
k_{3}}^{\eta}(\omega)-\Phi_{k_{1}k_{4},k_{2}k_{3}}^{\eta}(0)\Bigr]\delta%
^{\eta\eta^{\prime}}.
\label{W-omega}%
\ee
Here $\tilde V$ is the static part of the effective residual
quasiparticle interaction. In the present work, as in
Refs.~\cite{LRT.08,LRTL.09}, it is derived from the covariant energy
density functional (CEDF) with the parameter set NL3~\cite{NL3} as a
one-meson exchange interaction with a non-linear self-coupling
between the mesons. Pairing correlations are introduced into the
relativistic energy functional as an independently parameterized
term. In the 2q$\otimes$phonon version of the RQTBA the
energy-dependent residual interaction $\Phi(\omega)$ is derived from
the energy-dependent self-energy by the consistency condition and
calculated within the quasiparticle time blocking approximation.
This approximation means that, due to the time projection in the
integral part of the BSE, the two-body propagation through states
with a more complicated structure than 2q$\otimes$phonon is
blocked~\cite{Tse.07}.

In the present work we go a step further and consider a modified
version of the RQTBA which includes additional correlations between
two quasiparticles inside the 2q$\otimes$phonon configurations. This
version of the model has been proposed in Ref.~\cite{Tse.07} for the
non-relativistic case. It has been noticed that the energy-dependent
resonant part of the two-quasiparticle amplitude $\Phi (\omega)$ can
be factorized to extract the two-quasiparticle intermediate
propagator with the frequency shifted by the phonon energy. In the
relativistic RQTBA it takes the following form:
\be
\Phi^{\eta}_{k_1k_4,k_2k_3} (\omega) = %
\sum_{k_5k_6,\mu} \zeta^{\,\mu\eta}_{k_1k_2;k_5k_6}\,
\tilde{R}^{(0)\eta}_{k_5k_6} (\omega - \eta\,\Omega_{\mu})\,
\zeta^{\,\mu\eta *}_{\,k_3k_4;k_5k_6}\,, \label{phires} \ee
where $\tilde{R}^{(0)\eta}_{k_5k_6}(\omega-\eta\,\Omega_{\mu})$ are
the matrix elements of the two-quasiparticle propagator in the mean
field with the frequency shifted forward or backward by the phonon
energy $\Omega_{\mu}$,
\be \zeta^{\,\mu(+)}_{\,k_1k_2;k_5k_6} =
\delta^{\vphantom{(+)}}_{k_1k_5}\,\gamma^{(-)}_{\mu;k_6k_2}
-\gamma^{(+)}_{\mu;k_1k_5}\delta^{\vphantom{(+)}}_{k_6k_2} =
-\zeta^{\,\mu(-)\ast}_{\,k_2k_1;k_6k_5}\,, \ee
%
%
and $\gamma^{\eta}_{\mu;k_1k_2}$ are the quasiparticle-phonon
coupling vertices defined in Ref.~\cite{LRT.08}. In the graphic
expression of the amplitude (\ref{phires}) in the upper line of the
Fig.~\ref{f1} the propagator $\tilde{R}^{(0)\eta}_{k_1k_2}$ is
represented by the two straight nucleonic lines between the circles
denoting emission and absorption of the phonon by a single
quasiparticle with amplitudes $\gamma^{\eta}_{\mu;k_1k_2}$.
The extension of this model proposed in Ref.~\cite{Tse.07} consists
in the following: we introduce RQRPA correlations into the
intermediate two-quasiparticle propagator so that the two-phonon
configurations appear in the amplitude $\Phi (\omega)$ as it is shown
in the lower line of the Fig.~\ref{f1}. The analytic expression of
the new amplitude reads:
\be {\bar \Phi}^{\eta}_{k_1k_4,k_2k_3} (\omega) = \frac{1}{2}
\sum_{\mu,\nu}\frac{{\bar\zeta}^{\,\eta}_{\,\mu\nu;k_1k_2}\,
{\bar\zeta}^{\,\eta *}_{\,\mu\nu;k_3k_4}}
{\eta\,\omega-\Omega_{\mu}-\Omega_{\nu}}\,, \label{phiresc} \ee
where
\be {\bar\zeta}^{\,(+)}_{\,\mu\nu;k_1k_2} = \sum_{k_6} {\cal
R}^{(+)}_{\,\nu;k_1k_6}\,\gamma^{(-)}_{\,\mu;k_6k_2}\, - \sum_{k_5}
\gamma^{(+)}_{\,\mu;k_1k_5}\,{\cal R}^{(+)}_{\,\nu;k_5k_2} =
-{\bar\zeta}^{\,(-)\ast}_{\,\mu\nu;k_2k_1} , \ee
%
%
and ${\cal R}^{\eta}_{\,\nu;k_1k_2}$ are the matrix elements of the
RQRPA transition densities defined in Ref.~\cite{LRT.08} and
corresponding to grey circles together with two nucleonic lines in
Fig.~\ref{f1}. One can show that in the limit of the vanishing static
interaction $\tilde{V}$ between the two intermediate quasiparticles
Eq. (\ref{phiresc}) transforms to the Eq. (\ref{phires}) of the
original (R)QTBA.
As in conventional (R)QTBA, the elimination of double counting
effects in the phonon coupling is performed by the subtraction of
the static contribution of the amplitude ${\bar \Phi}$ from the
residual interaction in Eq. (\ref{W-omega}), since the parameters of
the CEDF have been adjusted to experimental data for ground states
and include therefore already essential phonon contributions to the
ground state. Therefore, the BSE in the two-phonon model has the
same form as Eq. (\ref{respdir}), but contains the amplitude
$\bar\Phi$ instead of $\Phi$. In order to take the
quasiparticle-phonon coupling into account in a consistent way, we
have first calculated the vertices $\gamma^{\eta}$ of this coupling
and the transition densities ${\cal R}^{\eta}$ within the
self-consistent RQRPA using the static residual interaction $\tilde
V$, as described in Ref.~\cite{LRT.08}. On the next step the BSE for
the correlated propagator $R^{(e)}(\omega)$
\be R^{(e)}(\omega) = {\tilde R}^{(0)}(\omega) + {\tilde
R}^{(0)}(\omega) \bigl[{\bar\Phi}(\omega) - {\bar\Phi}(0)\bigr]
R^{(e)}(\omega), \label{respdir2} \ee
has been solved in the Dirac-Hartree-BCS basis. Then, the BSE for
the full response function $R(\omega)$
\be R(\omega) = R^{(e)}(\omega) + R^{(e)}(\omega){\tilde V}R(\omega)
\ee
has been solved in the momentum-channel representations. The details
are given in Ref.~\cite{LRT.08}. To describe the observed spectrum
of a nucleus excited by a weak external field as, for instance, an
electromagnetic field $P$, the microscopic strength function $S(E)$
is computed as:
\be S(E) = -\frac{1}{2\pi}\lim\limits_{\Delta\to
+0}Im\Bigl[Tr(P^{\dagger}R(E+i\Delta)P) \Bigr]. \ee
%
%

\begin{figure*}[ptb]
\begin{center} \vspace{-2cm}
\includegraphics*[scale=0.8]{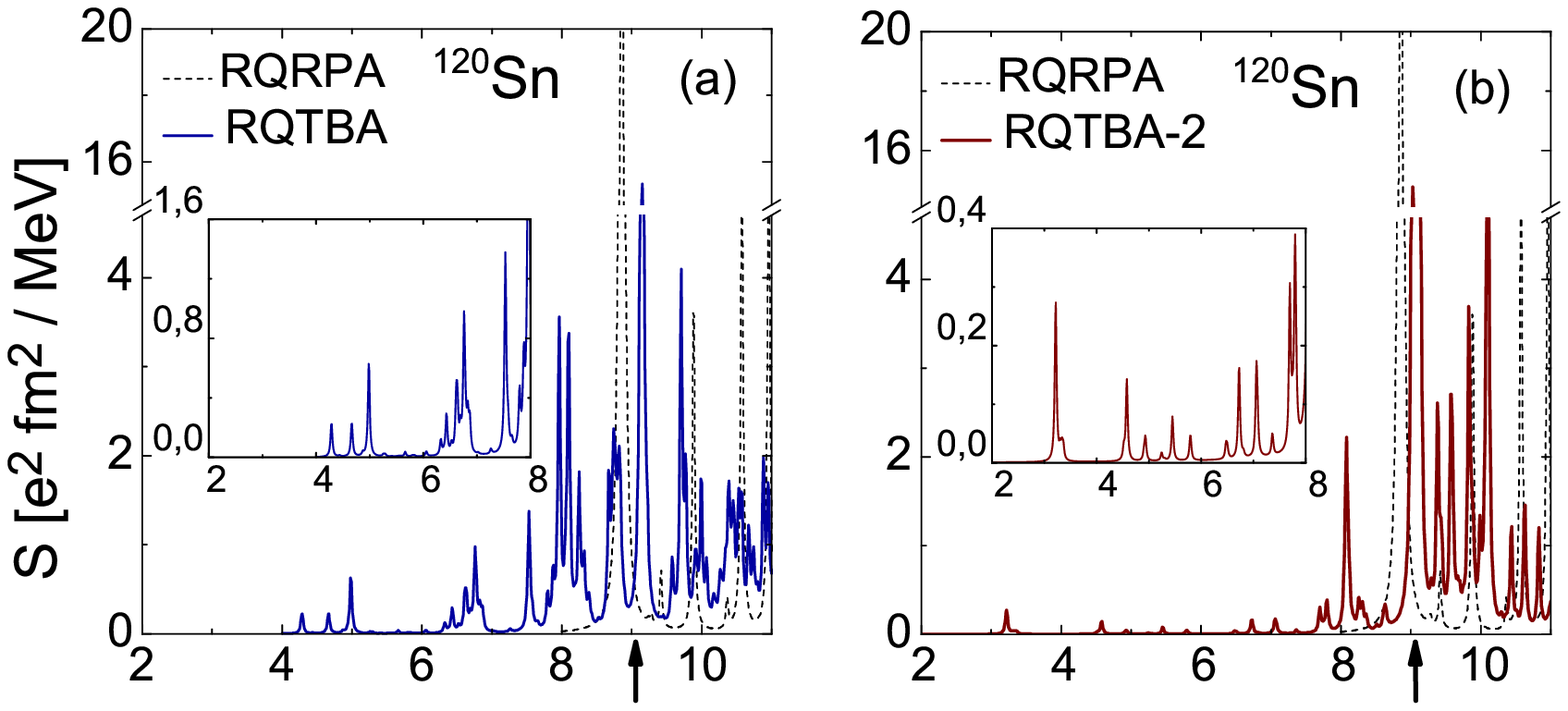}
\vspace{-2cm}
\end{center}
\begin{center}
\vspace{-3cm}
\includegraphics*[scale=0.8]{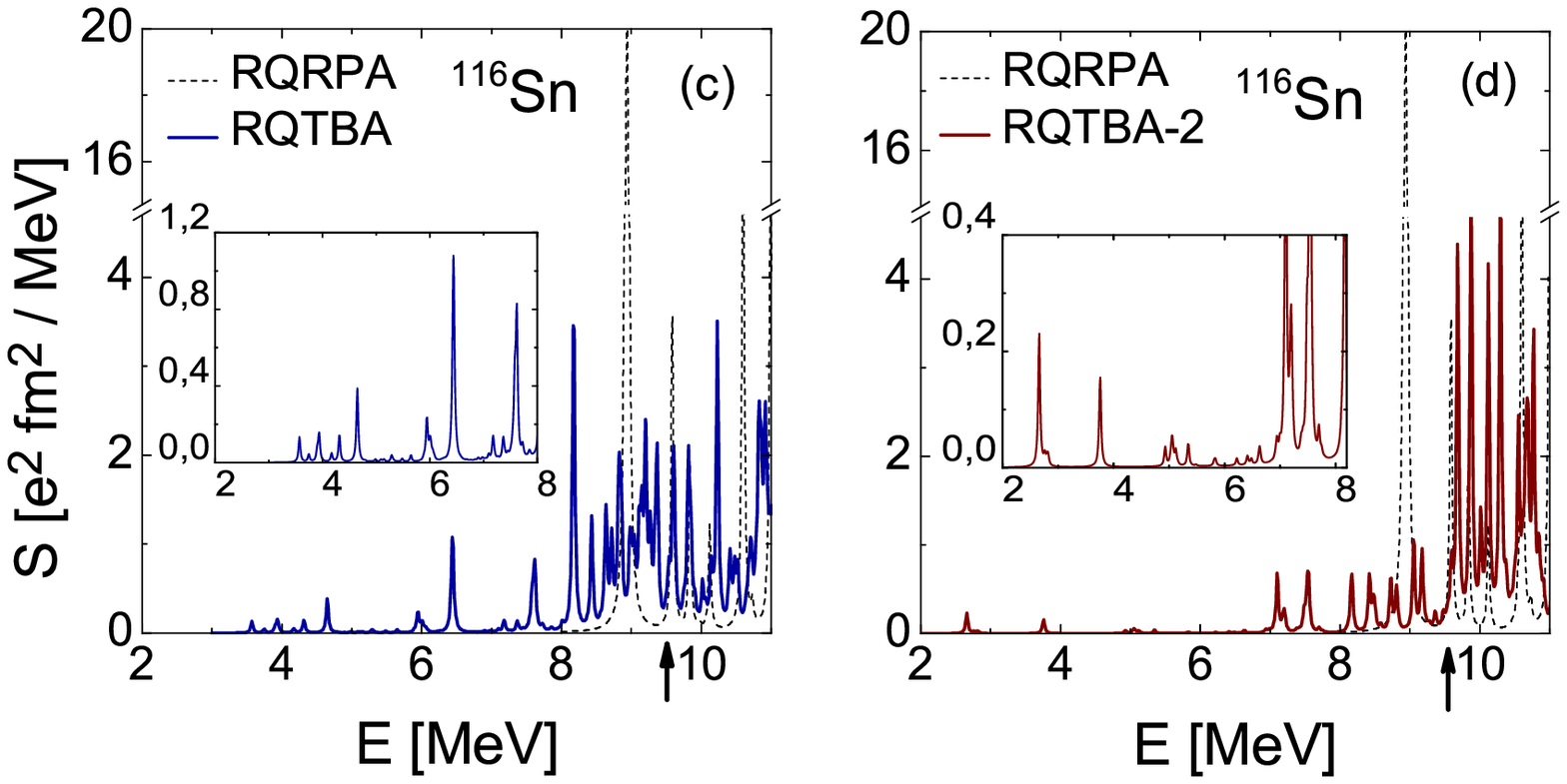}
\vspace{-1cm}
\end{center}
\caption{Low-lying dipole spectra of $^{116,120}$Sn calculated
within the RQRPA (dashed curves), RQTBA (blue solid curve, panels
(a,c)) and RQTBA-2 (red solid curve, panels (b,d)). A finite
smearing parameter $\Delta$ = 20 keV has been used in the
calculations. The inserts show the zoomed pictures of the spectra
below 8 MeV. The arrows indicate the neutron thresholds.}
\label{sn}%
\end{figure*}

To illustrate the effect of the additional static correlations on
spectra of nuclear excitations, we consider the dipole response of
tin and nickel isotopes in the area below the giant dipole resonance
(GDR). Fig.~\ref{sn} displays the dipole strength function in
$^{116,120}$Sn calculated within the conventional
RQTBA~\cite{LRT.08} and the two-phonon RQTBA-2 presented here. The
strength functions obtained in this way are compared with the
original RQRPA strength function because both of them originate from
RQRPA by different fragmentation mechanisms. The first observation
is that the total strength $\sum B(E1)\uparrow$ below the neutron
threshold is reduced in the RQTBA-2. For example, for $^{116}$Sn we
have 0.20 e$^2$ fm$^2$ below 8 MeV which agrees with 0.204(25) e$^2$
fm$^2$ obtained in the experiment of Ref.~\cite{Gov.98}. For
$^{120}$Sn, if we include the relatively strong state at 8.08 MeV
into the integration region, this quantity is 0.31 e$^2$ fm$^2$, in
agreement with the results of the quasiparticle phonon model (QPM)
of 0.289 e$^2$ fm$^2$~\cite{TLS.04}. One can also notice that in
both nuclei the major fraction of the RQRPA pygmy mode shown by the
dashed curve is pushed up above the neutron threshold by the RQTBA-2
correlations.


\begin{table}[ptb]
\caption{The energies, reduced transition probabilities and
anharmonicities of the lowest 1$^-$ states in $^{116,120}$Sn isotopes
calculated in the relativistic two-phonon model, compared to data of
Ref.~\cite{Bry.99}, see text for details.} \label{tab1}
\begin{center}
\tabcolsep=0.8em \renewcommand{\arraystretch}{1.0}%
\begin{tabular}
[c]{ccccc}\hline\hline
 &  & $\omega$(1$^-_1$) & B(E1)$\uparrow$ & $R_{\omega}$ \\
 &  & (MeV) & ($10^{-3}$ e$^2$ fm$^2$) &  \\
 \hline
 & RQTBA-2 & 2.66 & 14.5 & 0.94 \\
 $^{116}$Sn & Exp. & 3.33 & 6.55(65) & 0.94 \\
 \hline
 & RQTBA-2 & 3.22 & 16.9 & 0.95 \\
 $^{120}$Sn & Exp. & 3.28 & 7.60(51) & 0.92 \\
 \hline\hline
\end{tabular}
\end{center}
\vspace{-7mm}
\end{table}

The energies and the corresponding $B(E1)\uparrow$ values of the
lowest 1$^-$ states in the tin isotopes are listed in
Table~\ref{tab1}. The experimental energies, $B(E1)\uparrow$ values
and the $R_{\omega}$ values, $R_{\omega} =
\omega(1^-_1)/\bigl(\omega(2^+_1)+\omega(3^-_1)\bigr)$, are taken
from Ref.~\cite{Bry.99}. Notice, however, that the measurements with
the larger end point energies for the electron bremsstrahlung result
in the larger $B(E1)\uparrow$ values: 16.3(0.9) and 11.2(1.1) for
$^{116}$Sn and $^{120}$Sn, respectively~\cite{Gov.98,Oezel.phd}. A
good agreement with the data is obtained in spite of the fact that
these tiny structures at about 3 MeV originate by splitting-out from
the very strong RQRPA pygmy states located at the neutron thresholds,
that is described by the fully consistent inclusion of the two-phonon
correlations without any adjustment procedures.

In the RQTBA-2 the position of the first 1$^-$ state is basically
determined by the sum of the energies of the lowest 2$^+$ and 3$^-$
phonons. From the Eq. (\ref{phiresc}) one can see that the amplitude
$\bar{\Phi}(\omega)$ consists of the pole terms with the poles at the
energies which are sums of the two phonon energies. Therefore, the
energy of the first 1$^-$ state is approximately equal to the sum of
the energies of the lowest 2$^+$ and 3$^-$ phonons with some small
correction introduced by the static residual interaction $\tilde V$,
in agreement with data~\cite{Bry.99}. This correction is given by the
quantity $R_{\omega}$, whose deviation from unity characterizes the
two-phonon anharmonicity, see Table~\ref{tab1}. For example, in
$^{120}$Sn the energies of the $2^+_1$ and $3^-_1$ phonons calculated
within the RQRPA are obtained at 1.48 MeV and 1.90 MeV, respectively,
explaining the position of the 1$^-_1$ state at 3.22 MeV which is
identified experimentally as a member of the $2^+_1\otimes 3^-_1$
quintuplet.
%
%
\begin{figure*}[ptb]
\vspace{-1cm} \hspace{-2cm}
\includegraphics*[scale=0.9]{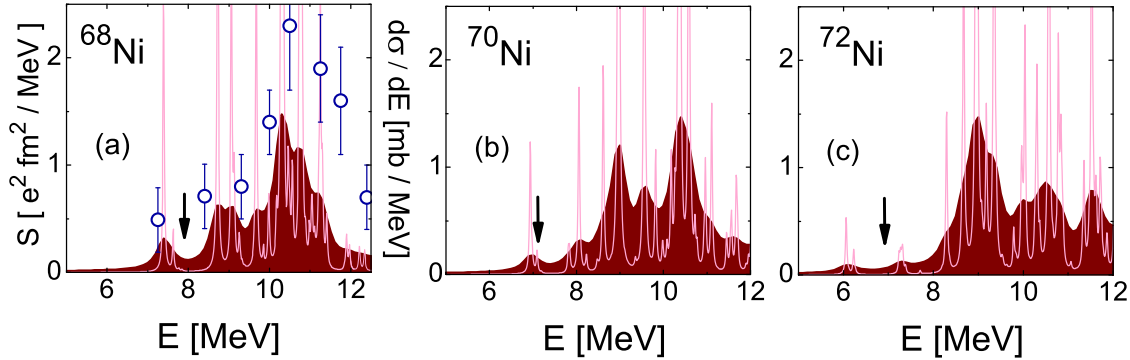}
\vspace{-5.8cm}
%
\caption{Low-lying dipole spectra of $^{68,70,72}$Ni calculated
within the RQTBA-2 with a smearing of 20 keV (thin curves, pink) and
200 keV (filled area). Panel (a) contains also the data from
Ref.~\cite{WBC.09} (open circles, units on the right). The arrows
show the neutron thresholds.}
\label{ni200}%
\end{figure*}

The electric dipole strengths in neutron rich $^{68,70,72}$Ni
isotopes are displayed in Fig.~\ref{ni200}. For all three isotopes
we found a redistribution of the low-lying strength as compared to
the RQTBA calculations of Ref.~\cite{LRTL.09}. The calculated
strength distribution in $^{68}$Ni has its maximum at 10.30 MeV and
the total strength below 12 MeV is 2.73 e$^2$ fm$^2$, while the
corresponding fraction of the energy weighted sum rule (EWSR) is
7.8\% of the total integrated photoabsorption cross section and 11\%
of the Thomas-Reiche-Kuhn sum rule. These characteristics are in
agreement with the recent data of Ref.~\cite{WBC.09}. The RQTBA-2
dipole strength distributions presented in Fig.~\ref{ni200} for the
$^{70,72}$Ni isotopes can be suggested as predictions for possible
future measurements in these nuclei.
%
%

In summary, the two-phonon version of the relativistic time blocking
approximation is presented. Within this model it has been shown how
the RQRPA modes are fragmented due to the coupling to two-phonon
configurations, thus explaining the physical connection between the
pygmy dipole mode and the 1$^-$ member of the $2^+_1\otimes 3^-_1$
quintuplet. A very reasonable description of the lowest 1$^-$ states
in $^{116,120}$Sn has been achieved within the fully consistent
scheme avoiding any adjustment procedures. The resulting low-lying
dipole spectra in $^{116,120}$Sn are compared with conventional
RQTBA calculations. It has been found that the two-phonon
correlations redistribute the fragmented strength as compared to the
2q$\otimes$phonon RQTBA so that the major fraction of the RQRPA
pygmy mode is pushed above the neutron threshold and therefore mixed
with the giant dipole resonance tail. The calculated low-energy
fraction of the electric dipole strength agrees also very well with
the available data for the tin isotopes and for the recently
investigated neutron-rich nucleus $^{68}$Ni. In general, the
relativistic two-phonon model presented here provides a new quality
of understanding of mode coupling mechanisms in nuclei. The method
is based on Green's function techniques and can be widely applied
also in other areas of quantum many-body physics.
%
%

Valuable discussions with H. Feldmeier and I. Mukha are gratefully
acknowledged. This work was supported by the Hessian LOEWE
initiative through the Helmholtz International Center for FAIR and
the Russian Federal Education Agency Program, project 2.1.1/4779.
\bigskip


\end{document}